\documentclass[journal=jacsat,manuscript=article]{achemso}

\usepackage[version=3]{mhchem} % Formula subscripts using \ce{}
\usepackage{xcolor}
\usepackage{subcaption}
\usepackage{tabularx}
\usepackage{caption}
\usepackage{pdfpages}

\usepackage{graphicx}% Include figure files
\usepackage{dcolumn}% Align table columns on decimal point
\usepackage{bm}% bold math
\usepackage{soul}
\usepackage{color}
\usepackage{xcolor}
\usepackage{amsmath}
\usepackage{amssymb}
\usepackage{amsmath}
\usepackage{float}
\usepackage{natmove}
\usepackage{multirow}
\usepackage{setspace}
\usepackage{array}
\usepackage{booktabs}
\usepackage{rotating}
\usepackage[normalem]{ulem}
\usepackage[colorlinks=true,linkcolor=blue,citecolor=blue,urlcolor=blue]{hyperref}
\usepackage{physics}
\usepackage{mathrsfs}
\usepackage[capitalise,noabbrev]{cleveref}
\usepackage[version=3]{mhchem} % Formula subscripts using \ce{}
\definecolor{Green}{rgb}{0.00,0.60,0.00}
\definecolor{Magenta}{rgb}{0.60,0.00,0.60}
\definecolor{BluBondi}{rgb}{0.00,0.58,0.71}
\definecolor{Orange}{rgb}{0.95,0.46,0.17}

\newcommand{\editor}[2]{%
  \expandafter\newcommand\csname #1note\endcsname[1]{%
    \textcolor{#2}{(\textbf{#1:} ##1)}}%
  \expandafter\newcommand\csname #1\endcsname[1]{%
    \textcolor{#2}{##1}}%
  \expandafter\newcommand\csname #1cancel\endcsname[1]{%
    \textcolor{#2}{\sout{##1}}}%
  \expandafter\newcommand\csname #1change\endcsname[2]{%
    \textcolor{#2}{\sout{##1} ##2}}%
  \newenvironment{#1text}{\color{#2}}{\color{black}}
}

\editor{DP}{BluBondi}
\editor{CAR}{red}
\editor{PR}{blue}
\editor{REV}{gray}
\newcommand{\drop}[1]{}

\author{Pavel Rukin}\affiliation{Cnr - Istituto Nanoscienze, via Campi 213/A, 41125 Modena (Italy)}
\author{Deborah Prezzi}\affiliation{Cnr - Istituto Nanoscienze, via Campi 213/A, 41125 Modena (Italy)}\email{deborah.prezzi@nano.cnr.it}
\author{Carlo Andrea Rozzi}\affiliation{Cnr - Istituto Nanoscienze, via Campi 213/A, 41125 Modena (Italy)}\email{carloandrea.rozzi@nano.cnr.it}
\title[]{Excited-state normal-modes analysis: the case of porphyrins}

\abbreviations{}

\begin{document}

%\begin{tocentry}
%
%\end{tocentry}

\begin{abstract}
Excited state normal modes analysis is systematically applied to investigate and compare relaxation and internal conversion dynamics of a free-base porphyrin with a novel functional porphyrin derivative. We discuss strenghts and limitation of the method, and employ it to predict very different dynamical behaviours in the two compounds and to clarify the role of high reorganization energy modes in driving the system towards critical regions of the potential energy landscape. For the functionalized porphyrin, we identify modes of vibrations along which the energy gap between different excited state potential energy surfaces within the Q band manifold may vanish, or be significantly reduced, with respect to the one observed in the bare porphyrin.
\end{abstract}

%%%%%%%%%%%%%%%%%%%%%%%%%%%%%%%%%%%%%%%%%%%%%%%%%%%%%%%%%%%%%%%%%%%
%%%%%%%%%%%%%%%%%%%%%%%%%%%%%%%%%%%%%%%%%%%%%%%%%%%%%%%%%%%%%%%%%%%
\section{Introduction}
%% Motivations
Studying the synthesis, photo-physics and photo-chemistry of porphyrin derivatives is a long-standing research topic, which gained particular attention in recent decades due to the compelling demand for improving solar energy harvesting devices \cite{TAKAGI2006104,Mandal2016,Hiroto2017,Paolesse17,Senge21,PANDA20122601,BISWAS2022112232}. In fact, these molecules, as well as chlorophyll (Chl) derivatives, act as reaction centers in both natural and artificial complex antenna systems \cite{Woller13, Otsuki18, Matsubara18, Auwarter2015, LLANSOLAPORTOLES2017296, GORKA2021102719, MASCOLI2020148156, CHEREPANOV2021112154}. 

Key to understand the initial steps of their photoexcited dynamics is the delicate interplay between the characteristic intense near-UV band (``$B$-band" or ``Soret band", around 400 nm) and the lower-energy visible band (``$Q$-band", in the range of 500-600 nm)\cite{Braun_1994_BP_THF}, which are qualitatively understood in terms of the Gouterman's four-orbital model \cite{GOUTERMAN1961}. 
Chemical functionalization of bare porphyrins (BP) generally preserves this excitation scheme, although it may affect the detailed shapes and positioning of the $Q$ and $B$ bands~\cite{Hiroto2017}. 

The dynamics between $B$ and $Q$ and within the $Q$ band of BP and some derivatives has been intensively studied with both theoretical and experimental methods~\cite{AKIMOTO1999177,Marcelli_jp710132s,KIM_acs.jpclett.5b01188,TDDFT_Ullrich, Falahati2018,Fortino_doi:10.1063/5.0039949}.
The internal conversion times of BP in benzene solution were estimated~\cite{AKIMOTO1999177} to be $40$~fs and $90$ fs for the $B\to Q_{y}$ and $Q_{y}\to Q_{x}$ transition respectively. In Ref.~\citenum{Marcelli_jp710132s}, it was reported that charge transfer states (CT) appearing in diprotonated porphyrin may favor $B\to Q_{y}$ internal conversion through the indirect $B\to CT$ step.
Time-resolved fluorescence experiments~\cite{KIM_acs.jpclett.5b01188} lead to the proposal of two different internal conversion pathways with different rates to explain $B$ band internal conversion in a tetra-phenyl-porphyrin, namely $B\to Q_{x}$ and $B\to Q_{y} \to Q_{x}$. In Ref.~\citenum{Falahati2018}, linear response TDDFT~\cite{TDDFT_Ullrich} and on-the-fly fewest switches surface hopping (FSSH)~\cite{Subotnik_2013} were also employed to explain the relaxation process between $B$ and $Q$ bands. It was shown therein that higher energy dark states (a band collectively called $N$) are involved into $B\to Q$ internal conversion, and that even $N \to Q_{x}$ population transfer is possible, even though less favorable than $N \to Q_{y}$, provided that enough excess of energy is available.
In Ref.~\citenum{Fortino_doi:10.1063/5.0039949}, the FSSH approach was applied to describe non-radiative relaxation processed within the $Q$-bands of chlorophylls showing the faster time crossing between the computed $Q_{x}$ and $Q_{y}$ population curves in the presence of the solvent as compared to the gas phase. 

In the experimental and theoretical studies above, the internal dynamics following photo-excitation, relaxation times and internal conversion pathways vary widely with respect to the ones of BP depending on the specific functionalization performed (for example tetraphenylporphyrin\cite{KIM_acs.jpclett.5b01188}, porphyrins bearing 0-4 meso-phenyl substituents \cite{Mandal2016}). This fact renders each functionalized system unique, and calls for theoretical characterization methods useful to find possible general trends.

Here we focus on a 5-Ethoxycarbonyl-10-mesityl-15-benzyloxycarbonyl porphyrin~\cite{Terazono15,Moretti2020} (FP). The synthesis of this molecule involves placing a carboxylic acid group directly on one or more of the BP meso-carbon atoms \cite{Terazono12}. This allows the construction of arrays in which the porphyrin macrocycles are close to each other and display an enhanced interaction with respect to, for example, the ones with hexa-phenylbenzene groups  \cite{Cho2006,Kodis06}. 
We perform an in-depth analysis of the active normal modes, including an investigation of the potential energy surfaces (PES) along their vibrations trajectories, and compare the results obtained for both  FP and BP. We show that excited-state normal-mode analysis, complemented by the calculation of per-mode reorganization energies and by a set of targeted scans along specific vibration modes, can unveil possible internal conversion pathways and point at specific regions of excited states PES that can be crucial in non-adiabatic dynamics.

%%%%%%%%%%%%%%%%%%%%%%%%%%%%%%%%%%%%%%%%%%%%%%%%%%%%%%%%%%%%%%%%%%%
\section{Methods}\label{sec:methods}

The analysis of normal modes on the excited state requires the calculation of per-mode reorganization energies (RE) and dimensionless Huang-Rhys (HR) factors, which provide a measure of the interaction strength between the electronic and vibrational states of the molecule. These quantities can be obtained within a displaced multi-mode harmonic oscillator model, which has been successfully applied in several works \cite{KRETOV20122143, Kretov13, Yurenevjp1031477, Zhigang_C3CS60319A, Rukin2015}. This approach is rigorously only valid as long as strong anharmonicities or Duschinsky effects \cite{Small1971} can be neglected. We will also work in Condon excitation regime and neglect spin-orbit effects, confining ourselves to the singlet manifold.

We are normally concerned about transitions (could be either optical absorption or internal conversion) occurring between an initial and a final electronic state, hereafter labeled $a$ and $b$, respectively. We aim at determining the influence of the vibrational modes calculated on a state $s$ on the $a \to b$ transition. The choice of the state $s$ for the calculation of the normal modes is usually dictated by the type of process under investigation. Often, the most meaningful choice is to make $s$ coincide with the final state $b$. Some other times the initial state or the ground state could be chosen as a cheaper approximation, in case a satisfying convergence on the excited state can not be achieve, even though vibronic replica in the absorption spectra will likely be less precise in this case.\cite{Wu_pyridine_modes}
Here we focus on the case where the initial state ($a$) geometry can be fully optimized, and only the gradient of the final excited state ($b$) is needed. Other cases are discussed in details in the Supporting Information.

\begin{figure}[H]
 \includegraphics[width=0.45\linewidth]{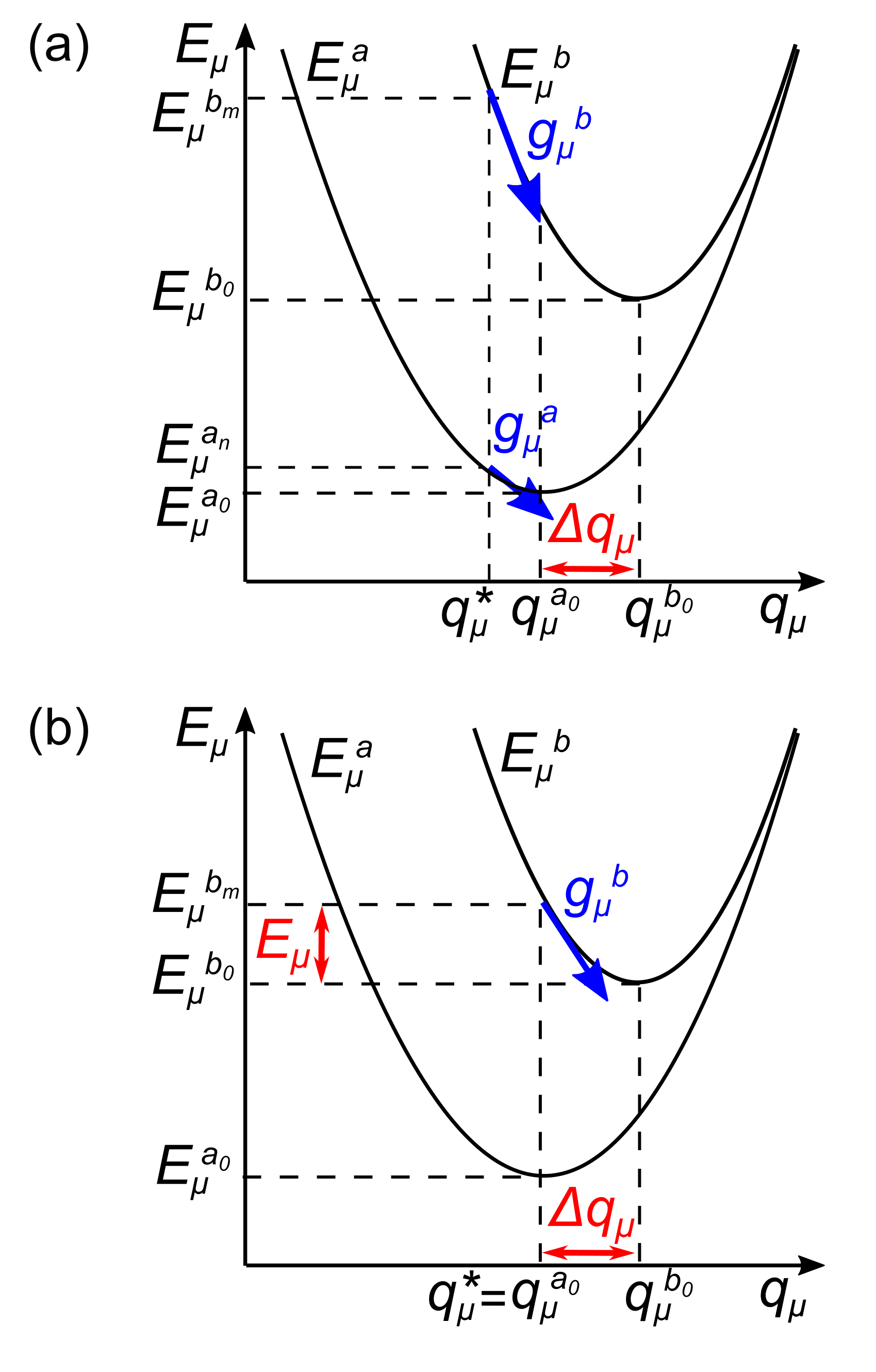}
 \caption{Definition of the shift $\Delta q_{\mu}$ along the mode $\mu$ in normal coordinates between an initial state $a$ and a final state $b$ in case of: (a) an arbitrary chosen normal coordinate $q_{\mu}^{*}$; (b) $q_{\mu}^{*}$ equals the normal coordinate at the minimum of state $a$.  $E_{\mu}^a$, $E_{\mu}^b$ are the PES along the mode $\mu$ of the states of interest; $q_{\mu}^{a_0}$, $q_{\mu}^{b_0}$  and $E_{\mu}^{a_0}$, $E_{\mu}^{b_0}$ are the coordinates and energies of the initial and final states minima, respectively; $E_{\mu}^{a_n}$, $E_{\mu}^{b_m}$ are the energies of the vibrational states; $g_{\mu}^{a}$,$g_{\mu}^{b}$ are the gradients of $a$ and $b$ PES at the $q_{\mu}^{*}$ normal coordinate.}
 \label{fig:Fig.M1}
\end{figure}

To define HRs and REs, we consider the shift of the potential energy surface (PES) $\Delta q_{\mu}$ between the initial state $a$ and the final state $b$ (see~\cref{fig:Fig.M1}a). The model represents the adiabatic harmonic potentials in the basis set of the normal modes {$\mu$} near the minimum of the PES of a selected state $s$.
Once the states of interest are identified, QM calculations are performed to obtained the equilibrium geometry in the initial state $a$. This geometry is then used to calculate the gradient of the total energy on the final state, $g^b$, at the coordinates corresponding to a vertical transition from $a$. The latter is obtained by computing the forces acting on each of the $N$ atoms of the system in the $b$ state. Then, the gradient $g^b$ is projected onto the normal modes of the state $s$. As such, we need to obtain the equilibrium geometry in the state $s$, for which the Hessian of the total energy is subsequently computed. From the diagonalization of the Hessian matrix, one can obtain the mode frequencies $\omega_{\mu s}$ and the reduced mass matrix $M$ [$(3N-6)\times(3N-6)$ matrix, whose  diagonal elements are $1/\sqrt{M_{\mu}}$ , where  ${M_\mu }$ are reduced masses, and non-diagonal ones are zeros], as well as the normalized transition matrix $L$ [$(3N)\times(3N-6)$] from normal $q$ [$(3N-6)$ vector] to Cartesian coordinates $X$ ($3N$ vector), where $X = L M q$.
Notably, the $L$ matrix is needed to compute the gradient operator projections onto the normal modes: 
\begin{equation}\label{eq:eq12}
  \hat g_\mu = \frac{\partial}{\partial q_{\mu}} 
  = \sum\limits_{i=1}^{N} \sum\limits_{j=1}^{3}
  \frac{\partial X_{ij}}{\partial q_{\mu}} 
  \frac{\partial }{\partial X_{ij}}
   = \sum\limits_{i=1}^{N}\sum\limits_{j=1}^{3} 
  L_{ij,\mu}^s
  \hat g_{ij},
\end{equation}
where $i$ runs over the $N$ atoms of the system and $j$ over the three Cartesian components.

Within the parallel harmonic approximation, the PES of a two-level system (such as in \cref{fig:Fig.M1}) for any normal mode coordinate $q_{\mu}^{*}$ can be written in terms of the displacements with respect to the PES minima as
\begin{equation}\label{eq:eq1}
  E^{a,b}(q^{*})=E^{a_0,b_0}+\frac{1}{2}\sum\limits_{\mu}M_{\mu}\omega_{\mu s}^2(q_{\mu}^{*}-q_{\mu }^{a_0,b_0})^2.
\end{equation}
Here, $E^{a_0}$ ($E^{b_0}$) and $q_{\mu a}^0$ ($q_{\mu b}^0$) are the energies and the normal coordinates of the initial (final)  state $a$ ($b$) at the minimum of its PES (see \cref{fig:Fig.M1}a); $\omega_{\mu s}$ and ${M_\mu }$ are the frequencies and reduced masses of the $s$ state.

\begin{figure}[H]
\centering
\includegraphics[width=0.5\linewidth]{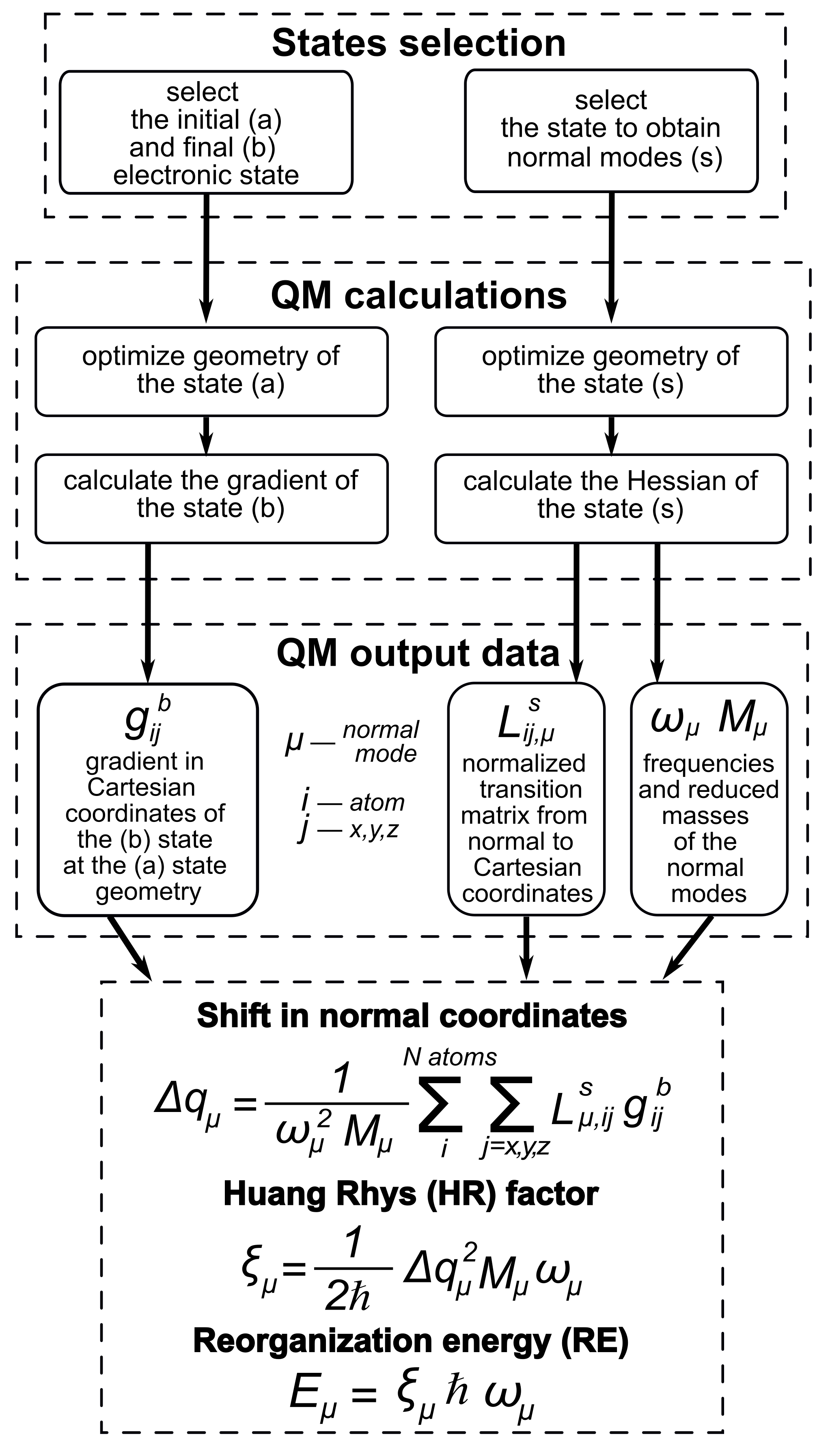}
\caption{Flowchart of the procedure used to compute HR factors and REs, as described in the text.}
\label{fig:Fig.M3}
\end{figure}

We can now obtain the projections of the gradients of the PES along the normal mode $\mu$ (see \cref{fig:Fig.M1}a) as
\begin{equation}\label{eq:eq2}
  g_{\mu}^{a,b} = \frac{\partial E^{a,b}(q^{*})}{\partial q_{\mu}^*}
  = M_{\mu}\omega_{\mu s}^2(q_{\mu}^{*} - q_{\mu }^{a_0,b_0}).
\end{equation}
Within the harmonic approximation, the difference of gradients $g_{\mu}^b-g_{\mu}^a$ does not depend on the initial point $q^{*}$. The normal coordinate displacement $\Delta q_{\mu}$ between the minima, the HR factor $\xi_{\mu}$, and the RE $E_{\mu}$ for each mode $\mu$ can be then written, respectively, as\cite{Lax52}
\begin{equation}\label{eq:eq4}
  \Delta q_{\mu}=q_{\mu }^{b_0}-q_{\mu }^{a_0}=\frac{-(g_{\mu}^{b}-g_{\mu}^{a})}{M_\mu \omega_{\mu s}^2 },
\end{equation}
\begin{equation}\label{eq:eq5}
  \xi_{\mu}=\frac{1}{2\hbar} M_{\mu} \Delta q_{\mu}^2  \omega_{\mu} =\frac{(g_{\mu}^{b}-g_{\mu}^{a})^2}{2 \hbar M_\mu \omega_{\mu s}^3}\,
\end{equation}
\begin{equation}\label{eq:eq6}
  E_{\mu}=\xi_{\mu} \hbar  \omega_{\mu} =\frac{(g_{\mu}^{b}-g_{\mu}^{a})^2}{2 M_\mu \omega_{\mu s}^2}.
\end{equation}
%%%%%%% 
As such, $q^{*}$ can be chosen in the most convenient way, e.g., such as to minimize computing time. If $q^{*}$ is taken as the minimum of the initial state $q_{\mu }^{a_0}$ (see \cref{fig:Fig.M1}b), coordinate differences, HR factors and REs assume the simplified form
\begin{equation}\label{eq:eq7}
  \Delta q_{\mu}=\frac{-g_{\mu}^{b}}{M_\mu \omega_{\mu s}^2 },
\end{equation}
\begin{equation}\label{eq:eq8}
  \xi_{\mu}=\frac{(g_{\mu}^{b})^2}{2 \hbar M_\mu \omega_{\mu s}^3},
\end{equation}
\begin{equation}\label{eq:eq9}
  E_{\mu}=\frac{(g_{\mu}^{b})^2}{2 M_\mu \omega_{\mu s}^2}. 
\end{equation}
Given $g^{a,b}_{\mu}$, we can then compute $\Delta q_{\mu}$, $\xi_{\mu}$, and $E_{\mu}$ for each normal mode $\mu$, according to \cref{eq:eq7,eq:eq8,eq:eq9}. The whole procedure for obtaining REs and HR factors from scratch by using quantum-chemical (QM) calculations is outlined in \cref{fig:Fig.M3}.

From the knowledge of HR factors, one can obtain the absorption spectrum with the inclusion of vibronic replicas (within the Franck-Condon approximation) by using the generating function approach \cite{Lax52,Yurenevjp1031477,KRETOV20122143,Kretov13,Rukin2015}, where the spectral line-shape $I(\omega)$ is defined in terms of the generation function $G(t)$ as follows:
\begin{equation}\label{eq:eq15}
  I(\omega)=\frac{1}{2\pi}
  \int\limits_{-\infty}^{\infty}
  G(t)e^{i \omega t} dt,
\end{equation}
\begin{equation}\label{eq:eq16}
  G(t) = exp\left[{\frac{it(E^{a_0}-E^{b_0})}{\hbar}}\right]
  \prod\limits_{\mu}
  exp \left[ -\xi_{\mu} \left(\coth{\frac{\hbar\omega_{\mu}}{2k_BT}}
  (1-\cos{\omega_{\mu}t}) - i\sin{\omega_{\mu}t}\right)\right]D(t),
\end{equation}
where $E^{a_0}-E^{b_0}$ is the purely electronic (zero-phonon) transition energy, $T$ is the temperature, and $\xi_{\mu}$ are the HR factors computed for the final state of the transition. The damping function is
\begin{equation}\label{eq:eq17}
  D(t)=e^{-\frac{\Gamma|t|}{\hbar}},
\end{equation}
where $\Gamma$ is the homogeneous line width \cite{Neese2009526}. In practice, since the high-frequency (``hard") modes define the structure of the spectrum, while the low-frequency ones (``soft") are responsible for the homogeneous broadening, $\Gamma$ can be calculated by splitting the normal modes into two groups and defining $\Gamma$ as the average FWHM of the soft modes \cite{FrankKamenetskii75,Rukin2015}
\begin{equation}
\Gamma=2\sqrt{2\ln{2}}\sigma, 
\end{equation}
where
\begin{equation}
\sigma^2=\sum\limits_{\mu={soft}}\xi_{\mu}\omega_{\mu}\coth{\frac{\hbar\omega_{\mu}}{2k_BT}}.
\end{equation}

HR factors and REs are especially useful to deepen the analysis of the coupling with the nuclear degrees of freedom, as normal modes characterized by large HR factors and REs are more likely to play a role in the electronic transition of interest. Indeed, within the harmonic approximation, REs give an estimate of the energy variation along the excited-state PES (see \cref{fig:Fig.M1}b). Once high-RE modes are identified, one can analyze the trajectories by moving the system along along those 'active' modes. This kind of analysis is not meant a substitute of explicitly dynamical methods~\cite{Rozzi_2018} as the time variable does not appear, however it provides a simplified, intuitive picture of the adiabatic PES landscape for individual modes.

The displacement along a mode $\mu$ ($\Delta q_{\mu}$) in Cartesian coordinates ($\Delta X_{\mu}$) can be obtained back from the normal coordinates by using the vector of $L$ matrix along the $\mu^{th}$ mode. Therefore the components of $\Delta X_{\mu}$ vector are
\begin{equation}\label{eq:eq13}
  \Delta X_{\ij,\mu}=\frac{1}{\sqrt{M_{\mu}}}L_{ij,\mu} \Delta q_{\mu}.
\end{equation}
Further, from starting configuration $X_{init}$, typically located at one PES bottom, deformed configurations following the $\mu$ mode can be computed as
\begin{equation}\label{eq:eq14}
  X_{\mathrm{final}}=X_{\mathrm{init}}+\Delta X_{\mu},
\end{equation}
where $X_{\mathrm{final}}$ indicates the vector of the displaced coordinates along the normal mode, while $X_{\mathrm{init}}$  is the vector of the initial coordinates (the minimum of the initial state $a$). At each desired $X_{final}$ vertical excitation energies can be computed, possibly point at "hot" points at which the PES of different electronic state get critically close to each other, or cross. An example of it can be found below in \cref{fig:fig.R5}.

%%%%%%%%%%%%%%%%%%%%%%%%%%%%%%%%%%%%%%%%%%%%%%%%%%%%%%%%%%%%%%%%%%%
\section{Results and discussion}

In this section we consider, side by side, BP and FP. We will apply the procedure detailed in the previous section to characterize the PES landscape for modes actively taking part to both $B\to Q$ and $Q\to Q$ internal conversion processes. The QM calculations to compute HR factors and REs are performed within the DFT and TDDFT frameworks by using the Gaussian16 package~\cite{g16}. The hybrid range-corrected CAM-B3LYP \cite{CAMB3LYP} functional, together with the 6-311(d,p) basis set, is used to determine both the equilibrium geometries and the gradients of the $a$, $b$ and $s$ states. The effect of the solvent (in our case tetrahydrofuran, THF) is included through the polarizable continuum model (PCM)~\cite{Tomasi05}. The optical transitions are characterized according to the natural transition orbital (NTO) analysis~\cite{NTO} of the TDDFT transition density as implemented in the Multywfn package~\cite{Multiwfn}.

\subsection{Vibronic effects in absorption spectra}
\begin{figure}[H]
\centering
\begin{subfigure}[H]{0.9\linewidth}
\centering
\includegraphics[width=\linewidth]{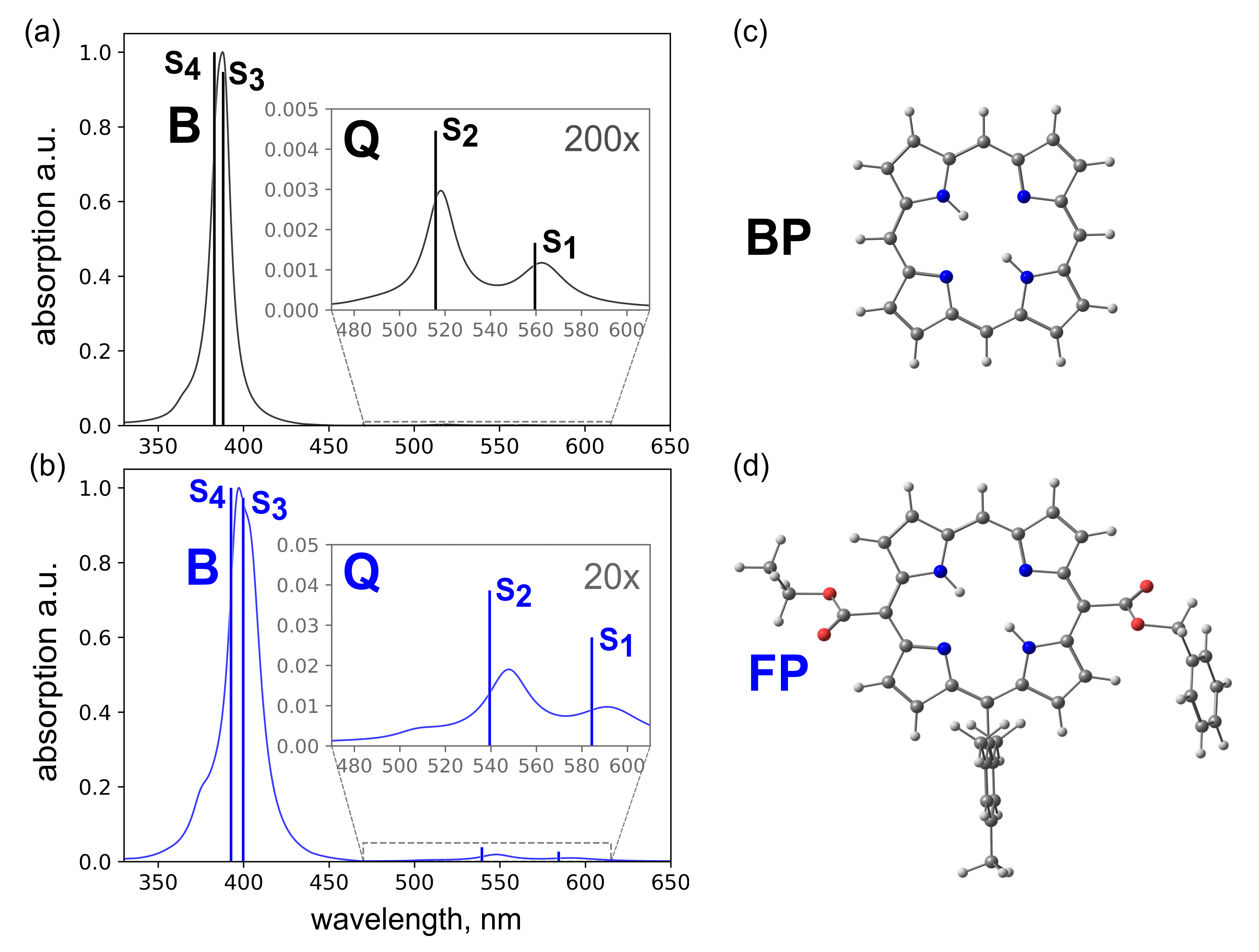}

\end{subfigure}
\caption{Calculated linear absorption spectra including vibronic effects and broadening of BP in black (panel a) and FP in blue (panel b). Vertical lines indicate vertical transition energies normalized to the maximum of the oscillator strength. Panel (c) and (d) show ground state structures of BP (black) and FP (blue).}
\label{fig:fig.R1}
\end{figure}

\cref{fig:fig.R1} shows the calculated absorption spectra of both BP (black curve, panel a) and FP (blue curve, panel b). Vertical lines indicate zero-phonon excitations as resulting from TDDFT simulations (see the Methods section). 
Here we follow the common convention and indicate the ground state as $S_0$. The (singlet) electronic excited states $S_1$ and $S_2$ belong to the Q band, while $S_3$ and $S_4$ to the B band. In particular, $S_1$ and $S_2$ correspond to the two non-degenerate $Q_{x}(0-0)$ and $Q_{y}(0-0)$ electronic excitations, arising from the lowered symmetry of BP with respect to metallo-porphyrins~\cite{Spellane_doi:10.1021/ic50204a021} ($D_{2h}$ {\it vs} $D_{4h}$) due to the presence of NH protons. As we will see later in the discussion, this lowered symmetry not only affects the spectra but also plays an important role for the internal conversion dynamics.

A direct comparison between the electronic excitations for the BP (a) and FP (b) shows that the excitation sequence remains the same upon functionalization, except for an overall redshift of the energies in the FP case, which are in quite good agreement with experimental data, as reported in \cref{tab.Tab1}. The strength of the Q band freatures is larger in FP than in BP, in agreement with experimental data~\cite{Braun_1994_BP_THF}. In addition, the electronic excitations are accompanied by two phonon replicas each ($Q_{x}(0-1)$,$Q_{y}(0-1)$), whose experimental values are also reported in \cref{tab.Tab1} (see discussion below).

\begin{table}[]
\centering
\setlength{\tabcolsep}{3pt}
\renewcommand{\arraystretch}{1.2}
\begin{tabular}{ c|c|c|c|c|c } 
 Band    & BP$_{exp}^{\cite{Braun_1994_BP_THF}}$ & BP$_{calc}$ & FP$_{exp}^{\cite{Terazono15}}$ & FP$_{calc}$ & State \\ 
 \hline
 Q$_x$(0-0) & 616  & 560      & 637  & 584     & S$_1$  \\ 
 Q$_x$(0-1) & 561  &          & 582  &         &   \\ 
 Q$_y$(0-0) & 518  & 519      & 543  & 540     & S$_2$  \\ 
 Q$_y$(0-1) & 487  &          & 506  &         &   \\
B        & 392  & 383/388  & 408  & 393/400 & S$_3$/S$_4$  \\
\hline
\end{tabular}
\caption{Comparison between experimental data\cite{Braun_1994_BP_THF, Terazono15} and calculated vertical transitions (in nm). The solvent is THF \cite{Braun_1994_BP_THF} and DCM \cite{Terazono15} for experimental data in BP and FP, respectively; given the similar dielectric constant, THF is used for both BP and FP in TDDFT calculations to ease the comparison.} 
\label{tab.Tab1}
\end{table}

%%%%%%%%%%%%%%%%%%%%%%%%%%%%%%%%%%%%%%%%%%%%%%%%%%%%%%%%%%%%%%%%%%%

\begin{figure}[H]
\centering
\includegraphics[width=0.98\textwidth]{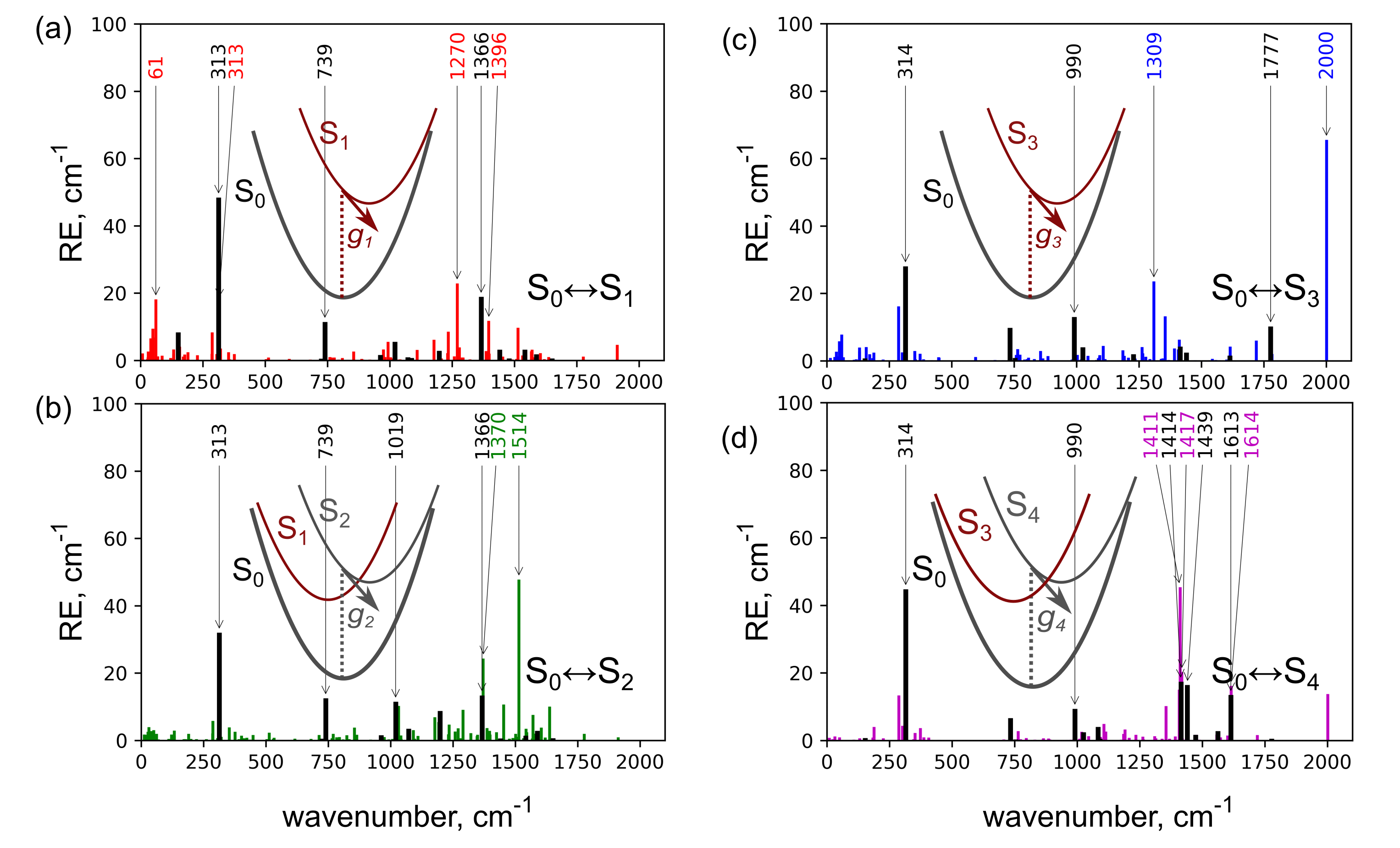}
\caption{Calculated per-mode reorganization energies (REs) based on transitions from the ground state ($S_0$) to (a) $S_1$ (b)  $S_2$  (Q band) and (c) $S_3$ (d) $S_4$  (B band) excited states . REs of FP modes are in colors (red for $S_1$, green for  $S_2$, blue for $S_3$, magenta for  $S_4$); BP REs are in black. The PES depicted in dark red in the schematic indicates the state chosen to compute the basis set of the normal modes for RE calculations.}
\label{fig:fig.R2}
\end{figure}

Starting from the purely electronic spectra (\cref{fig:fig.R1}, vertical bars), one can estimate the effect of molecular vibrations by computing the REs, as detailed in the Methods Section. In the following, we focus on vibrational modes with high REs (hereafter called {\it active modes}), which  are the ones contributing the most to the vibronic progression of the absorption spectrum and to the broadening of the peaks. 
\cref{fig:fig.R2} displays the per mode REs of BP (black bars) and FP (colored bars) for the transition from the ground to the different excited states, where the set of normal modes of $S_1$ was used to compute the REs for the Q band transitions (from $S_0$ to $S_1$ and $S_2$) and the $S_3$ set for the B band ones (from $S_0$ to $S_3$ and $S_4$). As can be noted by comparing the different panels of \cref{fig:fig.R2}, the REs show an overall increase and a more spread distribution upon functionalization, irrespective of the chosen transition. In fact, BP (black bars) shows quite sparse and rather few active modes. On the contrary, FP shows a more spread ``bath" of active vibrational modes, due to the further symmetry lowering caused by the presence of the functional groups external to the core ring.

The absorption spectra computed by including vibronic effects are reported in \cref{fig:fig.R1}. The vibration-induced homogeneous broadening, with $\Gamma$ obtained as the mean FWHM of soft modes below 400 cm$^{-1}$. The difference between BP and FP in RE values and their distribution here yields different values for $\Gamma$, i.e. 514, 378, 458, 380 cm$^{-1}$ for the first 4 transitions of FP and, correspondingly, 389, 295, 279, 352 cm$^{-1}$ for BP. In the high-frequency-mode range, the different REs between BP and FP give rise instead to different vibronic progressions. Specifically, in the case of BP, a shoulder to the B band appears at $\sim$370 nm, which can be attributed to the 1777 cm$^{-1}$ mode for the $S_0 \to S_3$ transition and to the set of modes in the range 1400-1600~$cm^{-1}$ for the $S_0 \to S_4$ transition. Vibronic replicas are instead almost negligible in the Q band due to absence of particularly high REs modes. 
Moving to FP, we find that the shoulder of the B band ( $\sim$378 nm) is noticeably more pronounced, and originates from the contribution of the 2000 cm$^{-1}$ mode for the $S_0 \to S_3$ transition and of the set of modes at 1400-1600~$cm^{-1}$ for the $S_0 \to S_4$ transition. In addition, a vibronic peak at $\sim$510 nm arises in the Q band, mostly due to a high-RE mode at 1514 cm$^{-1}$ for the $S_0 \to S_2$ transition. The position of the $S_2$ vibronic peak is close to the experimental Q$_y$(0-1) (\cref{tab.Tab1}), while it is impossible to clearly define a vibronic peak corresponding to Q$_x$(0-1) due to slightly blue shifted $S_1$ electron peak [Q$_x$(0-0)]. Indeed, as already shown for BP~\cite{Santoro2008,MINAEV2006308}, it is known that the inclusion of Herzberg–Teller effect is important to better reproduce absorption line shapes, which is however beyond the scope of this work.

\subsection{PES along active normal modes}

\begin{figure}[H]
\centering
\begin{subfigure}[b]{0.5\linewidth}
\centering
\includegraphics[width=\linewidth]{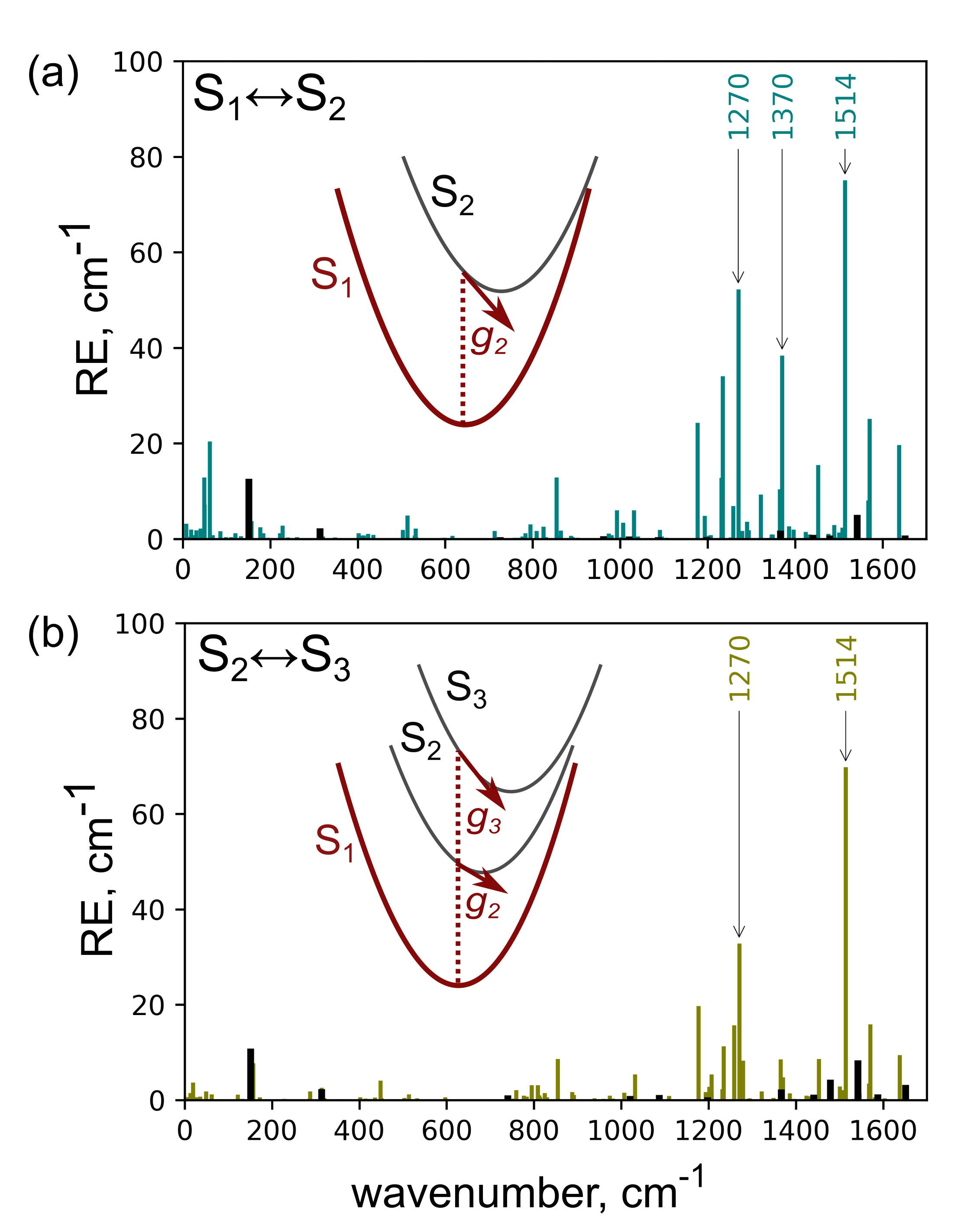}
\end{subfigure}
\caption{Calculated per-mode REs for (a) transitions from $S_1$ to $S_2$  (intra Q band) and (b) from $S_2$ to $S_3$ (from Q to B band) excited states . REs of FP modes are in colors (teal for $S_1$ to $S_2$, olive for $S_2$ to $S_3$; BP REs are in black. The PES depicted in dark red in the schematic indicates the state chosen to compute the basis set of the normal modes for RE calculations. Arrows and values highlight the most active modes.}
\label{fig:fig.R3}
\end{figure}

\begin{figure}[H]
\centering
\includegraphics[width=0.48\textwidth]{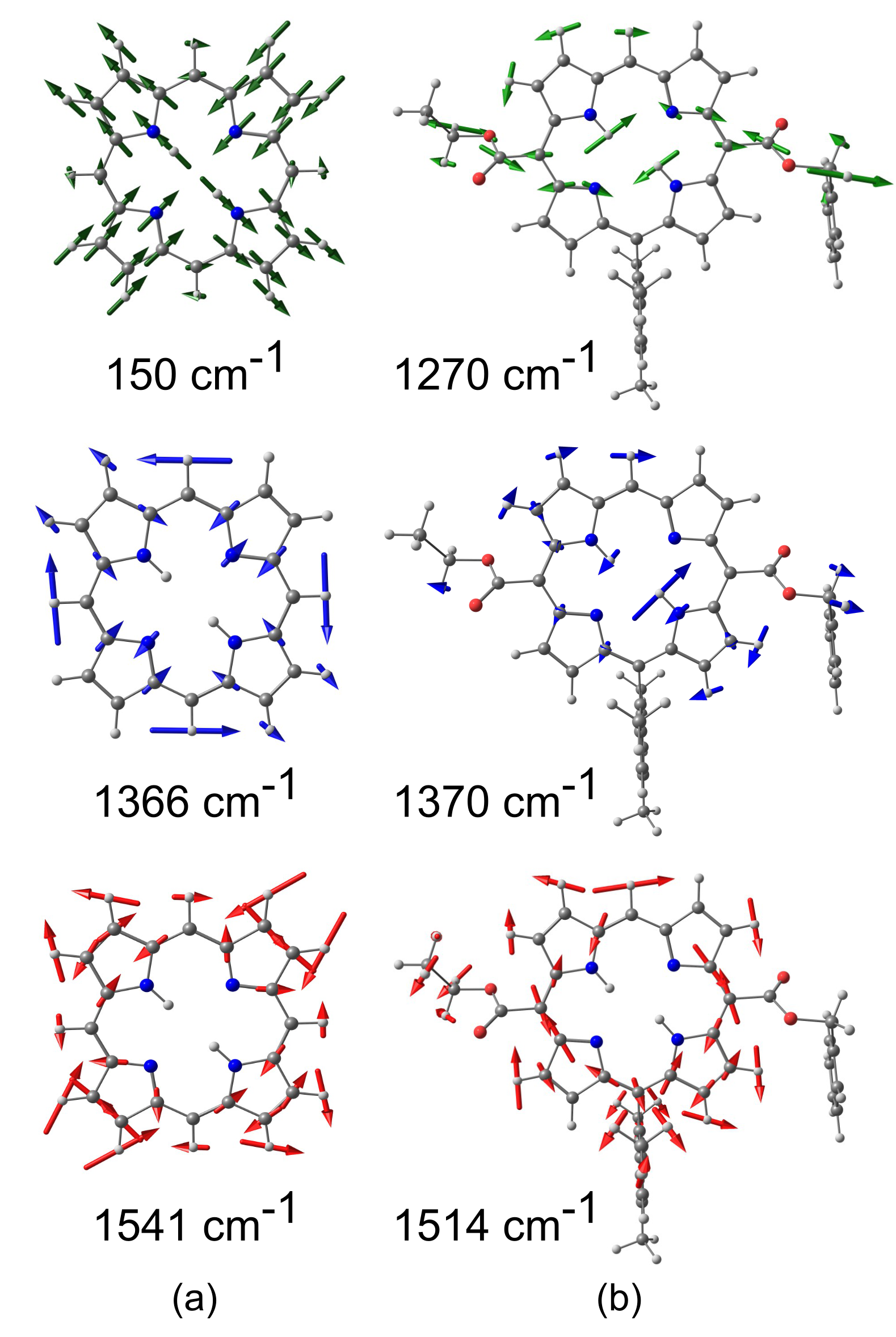}
\caption{Atomic displacementes for the normal modes of (a) BP (b) FP showing the higher REs in the $S_1 \to S_2$ transition.}
\label{fig:fig.R4}
\end{figure}

In addition to correcting the absorption spectra for vibronic effects, the per-mode REs are especially useful to understand the relaxation and internal conversion pathways. We have thus further examined the per-mode REs for different excited-state transitions, both within the Q band and between B and Q bands (see \cref{fig:fig.R3}, panel a and b, respectively). Here, REs are calculated on the set of $S_1$ modes. Again, a remarkable difference is found between BP (in black), with few sparse and weakly active modes, and FP (in colors), with several active modes, especially in the range above 1200 cm$^{-1}$. By focusing on the most active vibrations, (highlighted with arrows in \cref{fig:fig.R3}) and by inspecting the corresponding atomic displacements (see \cref{fig:fig.R4}), we find that these modes mostly differ in character for BP and FP, despite being all in-plane modes. In particular, the 1270 cm$^{-1}$ mode of FP involves large motion of the double carbon bonds of the ethoxy-carbonyl and benzyloxy-carbonyl connectors. Moreover both the mode at 1270 cm$^{-1}$ and at 1370 cm$^{-1}$ show asymmetric rocking of the N-H groups, not seen in the case of BP. Only the vibration at about 1500 cm$^{-1}$ has a similar pattern in the two molecules, although the amplitudes are less symmetric for FP. 
Notably, the latter mode has high RE for transitions both within the Q band and between Q and B bands, for both molecules. An in-depth analysis of the PES along this and other active modes by scanning vibrational trajectories can thus provide valuable information, detailed below.

\subsubsection{Single-mode analysis}

We have so far separately determined the electronic structure and the active vibrational modes on the excited states of BP and FP. Now we can merge this information by reconstructing the PES of the systems along the selected active modes.

\begin{figure}[H]
\centering
\includegraphics[width=\textwidth]{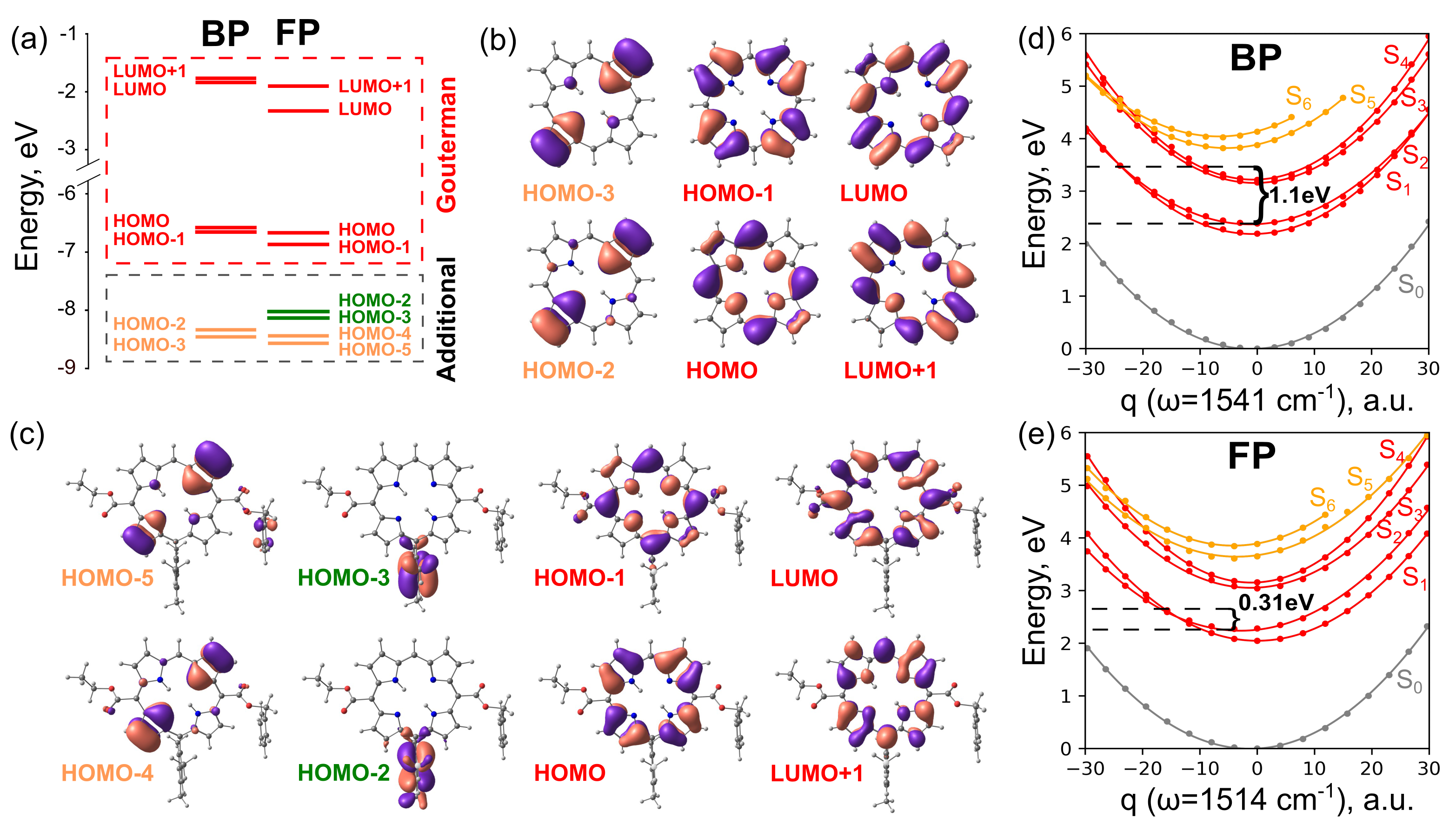}
\caption{(a) Energy-level scheme for BP and FP. (b, c) Selected molecular orbitals (MOs), contributing to the lowest excited states transitions, for BP and FP, respectively. (d, e) Trajectory scan along the modes at 1541~cm$^{-1}$ in BP and 1514~cm$^{-1}$ in FP, respectively. The $q=0$ coordinate corresponds to the optimized geometry in $S_1$, for which energy levels (a) and MOs (b,c)  are shown. The excited-state PES are colored according to the color code of the MOs (b,c) contributing the most to the transitions. 
The ground-state PES is in grey. Dots represent actual vertical energy calculations; solid lines are obtained by interpolation, according to the harmonic approximation.}
\label{fig:fig.R5}
\end{figure}

\cref{fig:fig.R5} shows the energy-level scheme (a) and the Kohn-Sham molecular orbitals (MOs) of the optimized $S_1$ state for BP (b) and FP (c). The HOMO and HOMO-1 orbitals look alike in the two molecules, except that the symmetries of BP HOMO-1 and HOMO ($A_u$ and $B_{1u}$ respectively are exchanged in FP, and the appearance of some density localized on the O atoms of the carboxylic acid group in FP. The LUMO and LUMO+1 MOs are instead remarkably affected by the connectors, which lower the symmetry and allow for a different mixing of the states. Moreover, by inspecting a few more states that are involved in higher-energy excitations (see \ref{fig:fig.R5}d-e and discussion below), we find that HOMO-2 and HOMO-3 in BP (\cref{fig:fig.R5}b) correspond to HOMO-4 and HOMO-5 in FP (\ref{fig:fig.R5}c); HOMO-2 and HOMO-3 in FP (\ref{fig:fig.R5}c) are instead completely localized on the mesityl group.

Starting from the above analysis, using formulas (\ref{eq:eq13}) and (\ref{eq:eq14}), we defined displaced geometries and computed the PES along the high-RE modes at 1541 cm$^{-1}$ in BP and at 1514 cm$^{-1}$ in FP, which were found to have similar characters in the two molecules (see \cref{fig:fig.R4}).

In \cref{fig:fig.R5} a cut of the PES along the two selected most active high frequency modes correspond to the $S_1$ to $S_2$ transition (see \cref{fig:fig.R3}a)  in both BP (panel d) and FP (panel e) is shown  The PES, calculated at the dotted points are interpolated according to the harmonic approximation; the PES color refers to the color code of the group of occupied MOs involved in the transitions (\cref{fig:fig.R5}a-c). This analysis allows one to understand whether the transition is dominated by a Gouterman ''dynamics'' (red) or other orbitals are involved (orange and green). Notably, there is no influence of the orbitals localized on mesityl group (green ones) on the B band, whereas they contribute in higher excited states (see Fig. S4). On the other hand, even though the scan only represents a specific section of the actual multidimensional space, it clearly shows that the $S_5$ and $S_6$ states (we label them collectively as $N$, as in Ref.~\cite{Falahati2018}) are crossing the B band. This is consistent with earlier suggested mechanism where upper energy levels favor $B\to Q_{y}$ internal conversion through the indirect $B \to N$ step~\cite{Falahati2018,Marcelli_jp710132s}. 

In addition to the crossing between $B$ and $N$ states, by looking at the reconstructed PES along the selected mode, we notice the existence of a point towards which the gap between the $S_1$ and $S_2$ PES tends to vanish. While this happens in both molecules, the energetics is rather different in the two cases. In fact the excess energy of the crossing point ($\bar{q}$) with respect to the $S_2$ minimum $\Delta E_{exc} = E_{S_2}(\bar{q}) - E_{S_2}(q_0)$ is 1.1~eV in the case of BP and 0.31~eV in FP. The same occurs with respect to the $S_1$ minimum. 

We conclude that the crossing point between the two Q-band states, driven by the mode at 1541~cm$^{-1}$ in BP (1514~cm$^{-1}$ FP), is much more easily accessible in FP than in BP. The fact that similar modes appear in both BP and FP, but with greatly enhanced REs in the latter, suggests the existence of a measurable effect in the intra-band dynamics, such a much faster internal conversion time.

\begin{figure}[H]
\centering
\includegraphics[width=1.00\linewidth]{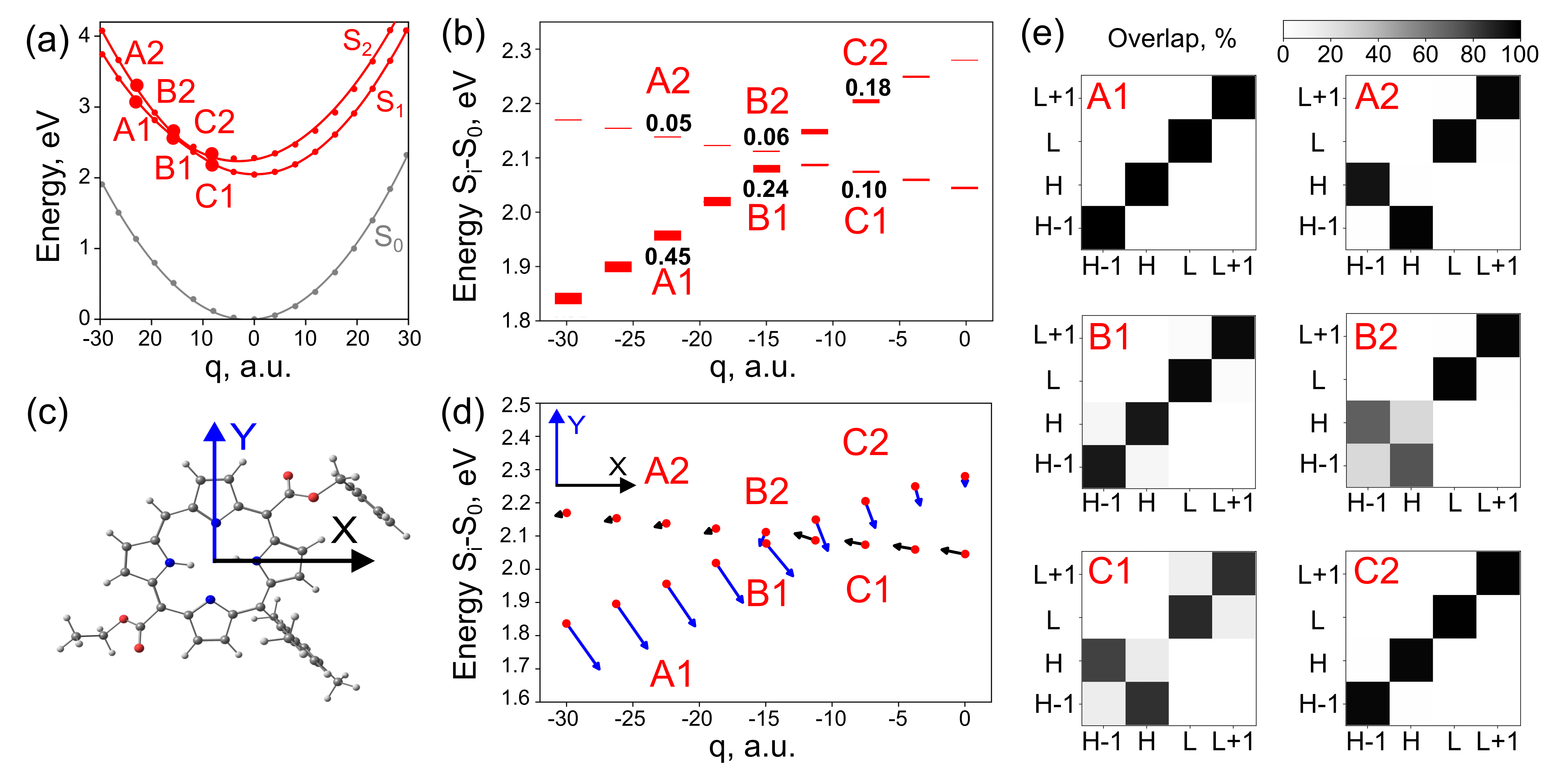}
\caption{(a) PES of the Q band of FP along the mode at 1514~cm$^{-1}$ ($S_0$ (grey), $S_1$,$S_2$(red)). Filled dots represent actual vertical energy calculations; solid lines are obtained by interpolation, according to the harmonic approximation. $A1$, $A2$, $B1$, $B2$, $C1$, $C2$ are selected transitions near the possible crossing, where numbers indicate the excited state order, i.e. $S_1$ or $S_2$.  
(b) Zoom near to the possible crossing displayed in (a).  Oscillator strengths are represented by the line thickness, while numerical values are indicated for the selected transitions. 
(d) Scan of the transition dipole moment (TDM) direction, with respect to the axes indicated in panel (c) (X - black, Y - blue), chosen as the symmetry directions of BP. The length of the arrows represents the TDM value. 
Energy values at the panel (b) and (d) are shown with respect to $S_0$ at each normal coordinate ($E_{S_i} - E_{S_0} $)
(e) NTO composition of the selected transitions (represented on the y axis) with respect to a reference state, here chosen to be $A1$ (x axis). Intensity of the color indicate the weights of $A1$ NTOs with respect to the Gouterman MOs. 
} 
\label{fig:fig.R6}
\end{figure}

Let us now examine more closely the Q band of FP (\cref{fig:fig.R6}a), by zooming in the region of the PES where the $S_1$ to $S_2$ gap vanishes (panels b, d) 
To identify the nature of the states on each surface around the zero-gap point, we considered three points along the trajectory centered around $\Delta q  \approx -15.7$, namely A, B, and C (\cref{fig:fig.R6}a), where numbers (1 or 2) indicates the order of the excited state, i.e. $S_1$ or $S_2$. For these selected q coordinates, we have analysed both the oscillator strength (panel b) and the transition dipole moments (TDM, panel d) of $S_1$ and $S_2$ along the symmetry directions of BP (x and y axes, as defined in panel c and corresponds to the states of $Q_x$ and $Q_y$ bands at \cref{tab.Tab1}). The oscillator strength, represented by the line thickness in panel b, decreases but remains non-zero moving from A1 to C2, while it remains closer to zero moving from A2 to C1. The TDM display a similar trend/behaviour, with x (black arrows) prevailing component (C2 → B1 → A1 ) or y (blue arrows) one (C1 → B2 → A2 ) again suggesting a crossing of the states.

For the same selected states, we also computed the NTOs in order to analyse their composition and character (see details in the Methods Section). Specifically, we used the NTOs of $S_1$ at point A (A1) -- here the Gouterman MOs -- as the basis set for our analysis, which is reported in \cref{fig:fig.R6}e. The grey scale indicates the composition of each selected state with respect to $A1$, partitioned onto its NTOs. A diagonal pattern indicates that the nature/character of the selected state is the same as A1; the presence of off-diagonal elements indicates instead that the nature of the state is different from that of the reference one. For instance, projecting the A2 state on A1 shows the exchange of HOMO and HOMO-1 orbitals, in accordance with the nature of Q$_x$ and Q$_y$ as described by the Gouterman model \cite{GOUTERMAN1961}.  As for oscillator strength and TDM trends, also the pattern found by the NTO analysis points to a crossing of the $S_1$ and $S_2$ states.

All of these analyses allow us to closely follow the character of the states around the zero-gap point of $S_1$ and $S_2$. This points to an actual exchange of the characters, instead of a repulsion of the PES. As the adiabatic approximation holds accurately far from this point it is therefore plausible to assume that the two surfaces will actually give raise to a conical intersection, or, at least, will become non-adiabatically coupled in the neighborhood of the crossing point. However the detailed geometry of the two surfaces and the classification of the intersection, requires dedicated methods, beyond the domain of the approximations we adopted here, and is left for future investigation.

\subsubsection{Two-mode/Coupled-mode analysis}
%\subsubsection{Two modes case}

\begin{figure}[H]
\centering
\includegraphics[width=1\linewidth]{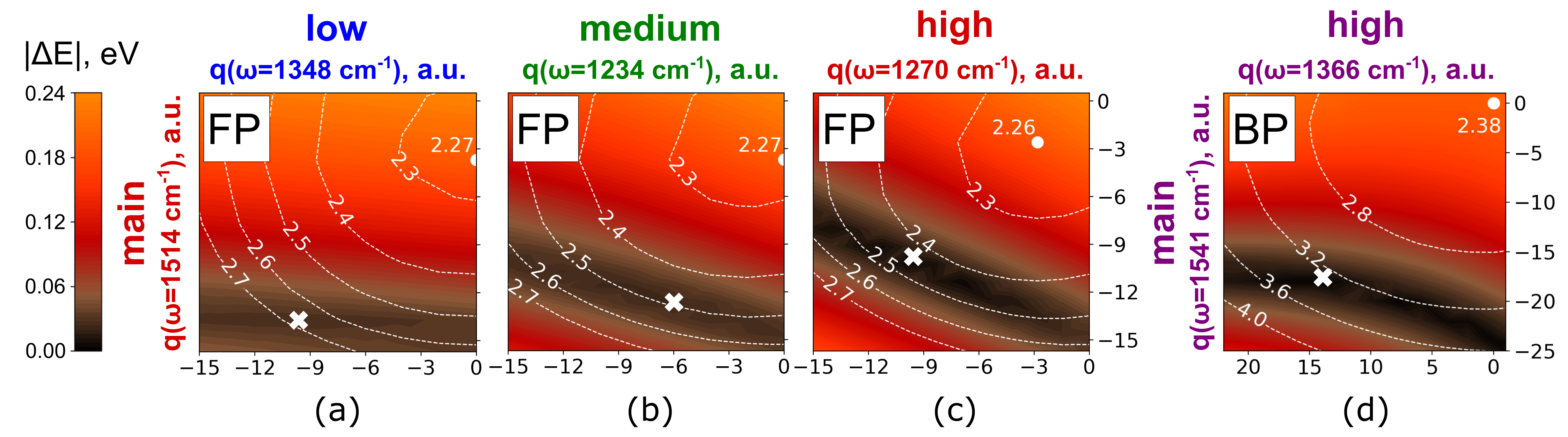}
\caption{Absolute value of the energy difference $|\Delta E|$ between $S_2$ and $S_1$ in FP (a-c) and BP (d), obtained by exploring the PES along two selected modes, i.e. the one with highest RE ({\it main}, y axis) and a second mode (x axis) chosen to have low (a), medium (b) or high (c,d) RE. The axes show the displacement along the modes in normal coordinates, where $q=0$ corresponds to the $S_1$ optimized geometry. White contour lines and values indicate the energy of $S_2$ PES relatively to the minimum of the $S_0$ PES. White circle shows (in eV) lowest energy of $S_2$ along considered modes. White cross indicate the point with the lowest $|\Delta E|$ value.}
\label{fig:fig.R7}
\end{figure}

In order to better understand the role of vibrations in the relaxation dynamics, we have explored the PES along additional modes (see \cref{fig:fig.R7}), which could influence the $S_2 \to S_1$ internal conversion by acting cooperatively with the highest-RE mode analysed previously. For FP, we explore the 2D space defined by the highest-RE mode ({\it main}, 1514 cm$^{-1}$, RE=75.08$cm^{-1}$) with three other modes in the same high-frequency region, having low (1348cm$^{-1}$, $RE=0.95~cm^{-1}$, panel a), medium (1234 cm$^{-1}$, $RE=34.08~cm^{-1}$ , b) and high (1270 cm$^{-1}$, $RE=52.29~cm^{-1}$, c) REs, respectively; for BP, we combine the highest-RE mode (1541 cm$^{-1}$, $RE=5.04~cm^{-1}$) with the next-highest-RE one (1366 cm$^{-1}$, $RE=1.79~cm^{-1}$ , d). The color maps reported in \cref{fig:fig.R7} display the energy difference $\Delta E$ between $S_2$ and $S_1$ in the 2D manifold defined by the two selected modes. This analysis provides valuable information, not only on the existence, location and shape of critical points/regions where the system can display strong non-adiabatic coupling (brown to black areas), but also on the energetic accessibility of these points, e.g.  $\Delta E_{exc}$ from the $S_2$ minimum in the touching point ($\bar{q}$). 

By looking at the different 2D maps computed for FP, we can notice that the weakly active mode at 1348~$cm^{-1}$ does not have any cooperative effect. In fact, the touching/critical region is almost parallel to the horizontal axis in the plot, that is, $\Delta E$ remains nearly the same by moving along the normal coordinate of this low-RE mode (\cref{fig:fig.R7}a). 
On the contrary, modes with higher RE values (panel b and c) can significantly modify the local landscape, leading to smaller $\Delta E$ at the same time with smaller excess energy from the $S_2$ minimum. The medium-RE mode (b panel) has $\Delta E_{exc}= 0.19$~eV at the crossing point on the PES, [$\bar{q} = (-6.0,-12.7,)$, marked with a cross]; the high-RE mode (c panel) leads to $\Delta E_{exc}= 0.16$~eV slightly further from the minimum [$\bar{q} = (-9.8,-10.2)$].

In conclusion, the comparison of the maps obtained for FP and BP (\cref{fig:fig.R7}c-d) clearly shows that the crossing region is much more accessible for FP than for BP, which confirms the possibility of a faster relaxation in FP, as anticipated from single-mode analysis. In fact, for BP we have found a barrier of 0.92 eV by [panel d, $\bar{q} = ( 14.0,-18.0)$], whereas the excess energy for FP given by the cooperative effect of the two highest-RE modes is four times lower, i.e.  $\Delta E_{exc}= 0.16$~eV.

%%%%%%%%%%%%%%%%%%%%%%%%%%%%%%%%%%%%%%%%%%%%%%%%%%%%%%%%%%%%%%%%%%%
\section{Conclusions}
We have employed excited state normal modes analysis to explore PES of BP and FP. This involved taking the following steps: 1) defining active modes (i.e. finding the particular set of modes of interest and selecting the modes with the highest RE values); 2) building transition energy scans along the active modes of interest; 3) analysing the states near the critical regions using the trends in changing oscillator strengths, the transition dipole moments (values and directions) and by comparing natural transition orbitals between the states of different structures along the normal modes scans. 

All of these analyses point at a crossing between the PES of $Q_x$ and $Q_y$ states and suggest that the considered functionalization of the porphyrin may substantially enhance the internal conversion within the $Q$ band. Moreover, the examination of the 2D PES along the active modes demonstrated that the FP has a much higher probability than the BP to reach the crossing point between the $Q_x$ and $Q_y$ states, upon $Q$ band excitation. The barrier between $Q_y$ minimum and the crossing point is just 0.16~eV for FP whereas it is 0.92~eV for BP.

We must bear in mind that the methodology followed here is bound to several constraints, as it is only rigorously valid in the adiabatic regime, within the assumption of harmonic PES for each of the considered vibrational modes and excluding vibrational mixing. However the method can be profitably exploited as a tool to quickly scan the PES for interesting points following the lead of the most active modes, at a reasonable computational cost. Therefore it can be used as a convenient initial step for exploring the effects of functionalization on the internal dynamics of porphyrins and other molecules. In the specific case we examined we have theoretically explained how a specific functionalization may have a huge impact on the internal conversion within the $Q$ band. We identified a particular vibrational mode responsible for driving the system into a conversion sweet-point, which opens the road on one side to a more advanced examination of the dynamics at or close the critical region, on the other to a more systematic study of different functionalization schemes.

\begin{acknowledgement}
This work was supported by Italian Ministry of University and Research, within the program PRIN 2017, grant no. 201795SBA3 - HARVEST. Computational time on the Marconi100 machine at CINECA was provided by the Italian ISCRA program. 
\end{acknowledgement}

%\begin{suppinfo}

%This will usually read something like: ``Experimental procedures and characterization data for all new compounds. The class will automatically add a sentence pointing to the information on-line:

%\end{suppinfo}

\providecommand{\latin}[1]{#1}
\makeatletter
\providecommand{\doi}
  {\begingroup\let\do\@makeother\dospecials
  \catcode`\{=1 \catcode`\}=2 \doi@aux}
\providecommand{\doi@aux}[1]{\endgroup\texttt{#1}}
\makeatother
\providecommand*\mcitethebibliography{\thebibliography}
\csname @ifundefined\endcsname{endmcitethebibliography}
  {\let\endmcitethebibliography\endthebibliography}{}

\newpage
\includepdf[pages=-]{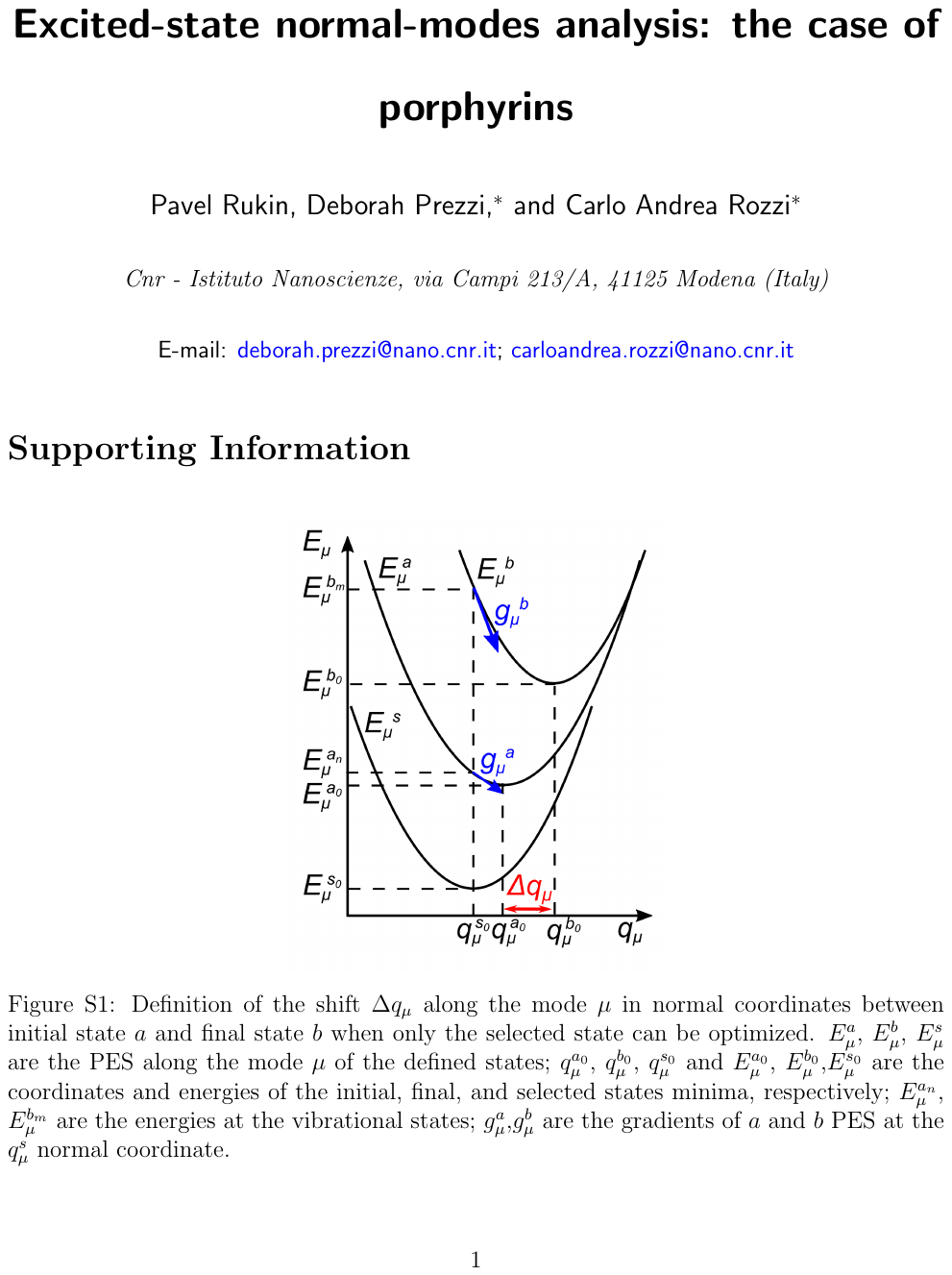}

\end{document}